\documentclass[twocolumn,prX,superscriptaddress]{revtex4}
\usepackage{times}
\usepackage[draft]{hyperref}
\usepackage{amsmath}
\usepackage{color}
\usepackage[applemac]{inputenc}
\usepackage{float}
\usepackage{graphicx}
 
\usepackage{color}
\newcommand{\tcr}{\textcolor{red}}

\begin{document}
\title{Observation of squeezed states with strong photon number oscillations}

\author{Moritz Mehmet}
\affiliation{Max-Planck-Institut f\"ur Gravitationsphysik (Albert-Einstein-Institut) and Institut f\"ur Gravitationsphysik, \\
Leibniz Universit\"at Hannover, Callinstr. 38, 30167 Hannover, Germany} 
\affiliation{Centre for Quantum Engineering and Space-Time Research - QUEST, Leibniz
Universit\"at Hannover,\\ Welfengarten 1, 30167 Hannover, Germany} 
\author{Henning Vahlbruch} \author{Nico Lastzka} \author{Karsten Danzmann} \author{Roman Schnabel\footnote{Roman.Schnabel@aei.mpg.de} }
\affiliation{Max-Planck-Institut f\"ur Gravitationsphysik (Albert-Einstein-Institut) and Institut f\"ur Gravitationsphysik, \\
Leibniz Universit\"at Hannover, Callinstr. 38, 30167 Hannover, Germany}

\begin{abstract}  
Squeezed states of light constitute an important nonclassical resource in the field of high-precision measurements, e.g. gravitational wave detection, as well as in the field of quantum information, e.g. for teleportation, quantum cryptography, and distribution of entanglement in quantum computation networks. Strong squeezing in combination with high purity, high bandwidth and high spatial mode quality is desirable in order to achieve significantly improved performances contrasting any classical protocols. Here we report on the observation of the strongest squeezing to date of 11.5\,dB, together with unprecedented high state purity corresponding to a vacuum contribution of less than 5\%, and a squeezing bandwidth of about 170\,MHz. The analysis of our squeezed states reveals a significant production of higher-order pairs of quantum-correlated photons, and the existence of strong photon number oscillations. 
\end{abstract}


\maketitle
\section{Introduction}
Squeezed states as well as number states (Fock states) are so-called nonclassical states. They allow the measurement, the communication and the processing of information in a way not possible with coherent states that are governed by vacuum fluctuations. Squeezed states of light and correlated photon pairs have been used in order to realize interferometric measurements with sensitivities beyond the photon counting noise \cite{XWK87,GSYL87,MSMBL02,WPAUGZ04,VCHFDS05,Goda08NatPhys}, to demonstrate the Einstein-Podolski-Rosen paradox \cite{EPR35,Aspect81,Ou92,Wagner08}, as a resource for quantum teleportation \cite{Bouwm97,FSBFKP98,BTBSRBSL03} and for the generation of Schr\"odinger kitten states for quantum information networks \cite{Ourjo06,Neerg06}. In view of applications in long distance quantum communication, purification and distillation of entangled photon number states and entangled two-mode squeezed states were experimentally demonstrated \cite{Ursin06NatPhys,DLHMFLA08,HSDFFS08}.\\
Fock states are characterized by photon counting detectors, whereas squeezed states are traditionally
characterized in the phase space of position and momentum-like operators \cite{Walls83,Breit97}. Appropriate
operators are the non-commuting amplitude and phase quadratures of the field, $\hat X_1$ and $\hat X_2$, respectively.
Their variances obey a Heisenberg uncertainty relation, $\Delta^2 \hat X_1 \cdot \Delta^2 \hat X_2 \geq
1/16$, with the vacuum noise variance normalized to 1/4. A squeezed state is realized if the quantum noise of
one quadrature is \emph{squeezed} below the vacuum noise level, at the expense of increased noise of the
other quadrature. For an overview we refer to Ref.~\cite{GerryKnight}. 
The relation between the quadrature operators and the photon number operator $\hat n$ is revealed by the Hamilton operator of the quantized harmonic oscillator:
\begin{equation}
    \hat H = \hbar \omega (\hat n + 1/2) = \hbar \omega (\hat X_1^2 + \hat X_2^2)\,,
    \label{hamiltonian}
\end{equation}
where $\hbar$ is the reduced Planck constant and $\omega$ the light's angular frequency. One can think of the two expressions in equation~(\ref{hamiltonian}) corresponding to the particle picture and the wave picture of quantum physics respectively. These can be used to calculate the mean photon number of squeezed states. Note, that this number is always greater than zero. Consider now a squeezed state without displacement in phase-space. Such a state is called a squeezed \emph{vacuum} state, and one can write $\langle \hat X_i^2 \rangle = \Delta^2 \hat X_i$. If such a state is pure, i.e. has minimum uncertainty, only \textit{even} photon numbers can be observed \cite{Yue76, DMM94}. Furthermore, if the squeezing effect is strong a so-called odd-even \emph{photon number oscillation} is realized, that includes not just the existence of photon pairs but also the existence of higher even-numbers of quantum-correlated photons.\\
This article presents the generation and characterization of squeezed vacuum states with strong odd-even photon number oscillations revealing a significant contribution of correlated photon numbers up to 10 and higher. This result is  made possible because our squeezed vacuum states show unprecedented quality, i.e. are strongly squeezed, are of high purity, and are in a well-defined spatial Gaussian fundamental mode. These properties are of great importance for applications in high-precision measurements as well as in quantum information. To the best of our knowledge, our squeezed light source also shows the broadest squeezing bandwidth ever observed and thus constitutes a bright source of quantum-correlated photons. 
%
\section{Experiment} 
\begin{figure}
\begin{center}
\includegraphics[width=0.49\textwidth]{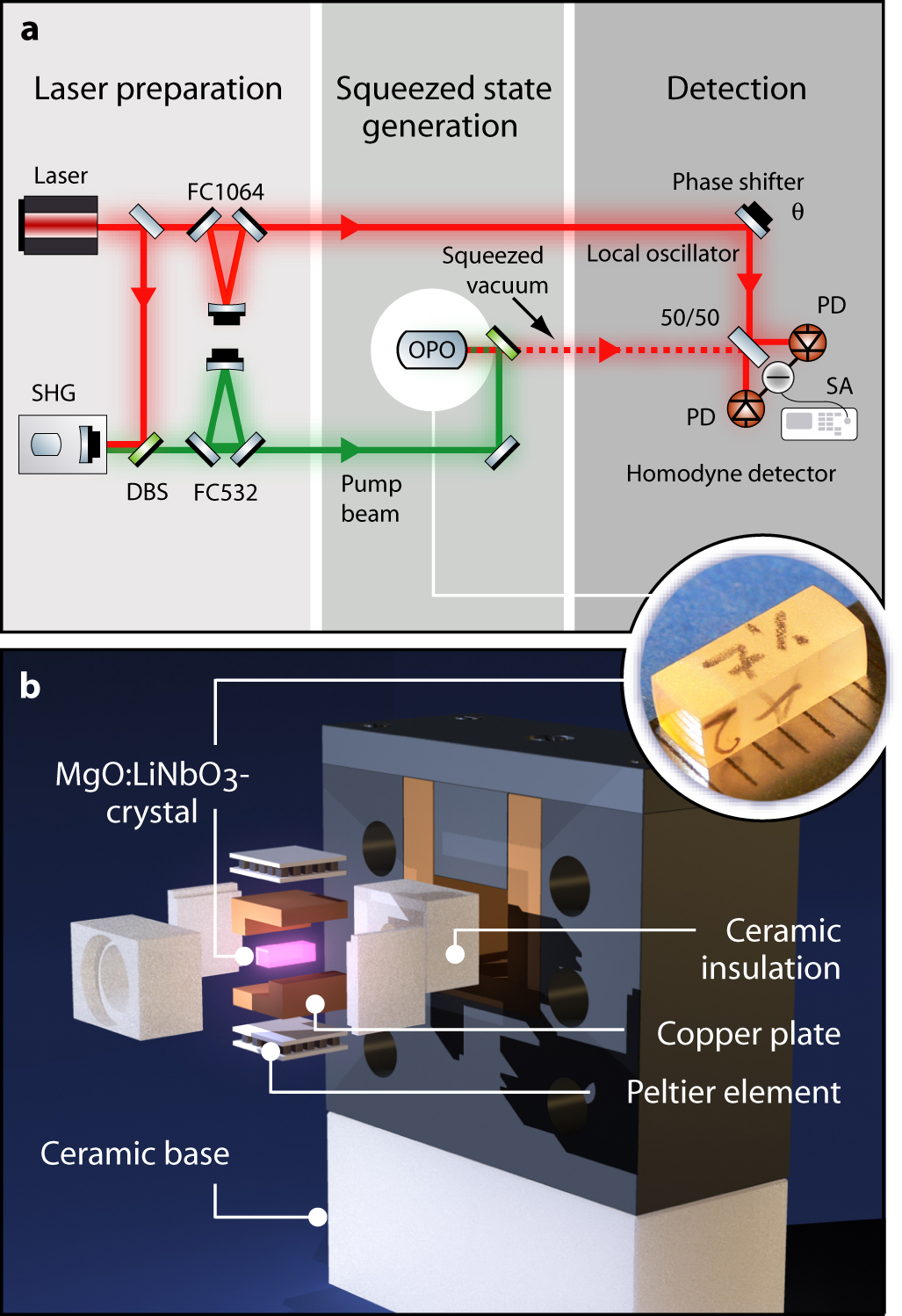} 
\end{center}
\caption{Experimental apparatus. {\bf{(a)}},
Schematic diagram of the experimental setup. A small part of the 1064\,nm output of the 2\,W laser system was used as local oscillator while the majority was sent to an external second harmonic generation (SHG) cavity which provided the 532\,nm pump beam to drive the OPO. Travelling wave filter cavities (FC1064 and FC532) in both paths were used to provide stable and well defined laser beams. Squeezed vacuum states at 1064\,nm were generated by type\,I optical parametric oscillation (OPO) below threshold in a nonlinear monolithic cavity.  A balanced homodyne detector was used to measure the states' quadrature variances. DBS: dichroic beam splitter; PD: photo diode; SA: spectrum analyzer. {\bf{(b)}}, Exploded assembly drawing of the oven enclosing the squeezed light source. The nonlinear crystal,
copper plates, peltier elements and thermal insulations are shown. The inlay shows a photograph of the monolithic
squeezed light source made from 7\,\% doped MgO:LiNbO$_{3}$. 
}
\label{experiment}
\end{figure}

At present the most successful squeezed light sources are based on parametric processes  
in nonlinear optical materials. In the regime of pulsed laser radiation the optical Kerr effect has recently enabled the observation of nearly 7\,dB squeezing \cite{DHCDAL08}. In the regime of continuous-wave laser radiation, squeezing of vacuum fluctuations by up to 10\,dB was observed utilizing optical parametric oscillation (OPO) below threshold \cite{Taken07,Vahlb08}.\\ 
The squeezed vacuum states investigated here were produced via type I OPO below threshold in a nonlinear standing wave cavity, similar to the experimental set-up in \cite{Vahlb08}. In order to observe high levels of squeezing the entire experiment was optimized for low phase fluctuations and low optical loss. A schematic diagram of the experimental setup is shown in Fig.\ref{experiment}(a). 
The main light source was a 2 Watt continuous wave laser at 1064\,nm. 
It was used to generate a second harmonic light field to pump the parametric squeezed light source and to provide the local oscillator (LO) beam for balanced homodyne detection. Both laser fields were sent through travelling wave filter cavities that were kept on resonance using Pound-Drever-Hall modulation-demodulation control schemes. The filter cavities therefore acted as optical low-pass filters that reduced potential phase noise and provided clean gaussian fundamental modes. The filtered second-harmonic pump beam had a power of up to 600\,mW and was subsequently injected into the OPO cavity.\\
In order to minimize the internal optical loss of the squeezed light source we designed the OPO as a monolithic (bi-convex) cavity, similar to the resonator used in \cite{Breitenbach95}. It was made from a 7\% doped Mg:O LiNbO$_3$ crystal measuring 6.5$\times$2$\times$2{.}5\,mm$^3$ in dimensions with end faces of 8\,mm radii of curvature (see inlay in Fig.~\ref{experiment}). While the back of the crystal had a high reflection coating for both the wavelengths involved, we chose a rather low power reflectivity of 88\,\% at 1064\,nm (and $<$\,1\,\% at 532\,nm) for the coupling front face in order to increase the ratio of out-coupling rate and cavity round trip loss. The crystal was controlled to keep its phase matching temperature close to 60\textdegree C using peltier elements, and the main laser frequency was tuned to match the resonance condition of the OPO cavity. An assembly drawing of the crystal mounting is shown in Fig. \ref{experiment}(b).\\ 
No control fields at or near the fundamental wavelength of 1064\,nm were injected into the squeezed light source during data taking. Thus all photons around 1064\,nm leaving the OPO cavity originated from the parametric down-conversion process. The out-coupled squeezed states, i.e. the photons at 1064\,nm, were separated from the 532\,nm pump upon reflection at a dichroic beam splitter (DBS) and subsequently analyzed by means of balanced homodyne detection. To this end the bright and spatially filtered LO beam was precisely adjusted to the spatial mode of the squeezed beam. We achieved a fringe visibility of 99{.}8\,\% at the homodyne detector's 50/50 beam splitter, which proves that the squeezed states were indeed in a well-defined gaussian fundamental mode.\\ 
A homodyne detector is the only known means to observe and characterize strongly squeezed states. Todays single photon detectors do not reach the same high quantum efficiencies and have an insufficient photon number resolution. It has been shown that homodyne detection data of quantum states can indeed provide their \textit{full} quantum information \cite{Bertrand87,Vogel89,Schiller96}. The phase angle $\theta$ between the local oscillator (LO) and the signal beam defines an arbitrary quadrature of the signal's quantum state $X(\theta)=\cos(\theta) X_{1} + \sin (\theta) X_{2}$ \cite{GerryKnight}. In our experiment $\theta$ was controlled with a piezo actuated mirror in the local oscillator path which allowed us to precisely set the phase angle. For example, to measure the squeezed quadrature $X(0)=X_1$ or the anti-squeezed quadrature $X(\pi/2)=X_2$. The collected data was readily converted into variances by a spectrum analyzer that was also used to select a certain Fourier frequency $f$ and resolution bandwidth $\Delta f$. The vacuum noise reference was measured with the same settings but with the squeezed field input blocked. 
\subsection{Observation of high-purity strongly squeezed states}\label{subsection1}
\begin{figure}
     \centerline{\includegraphics[width=0.5\textwidth]{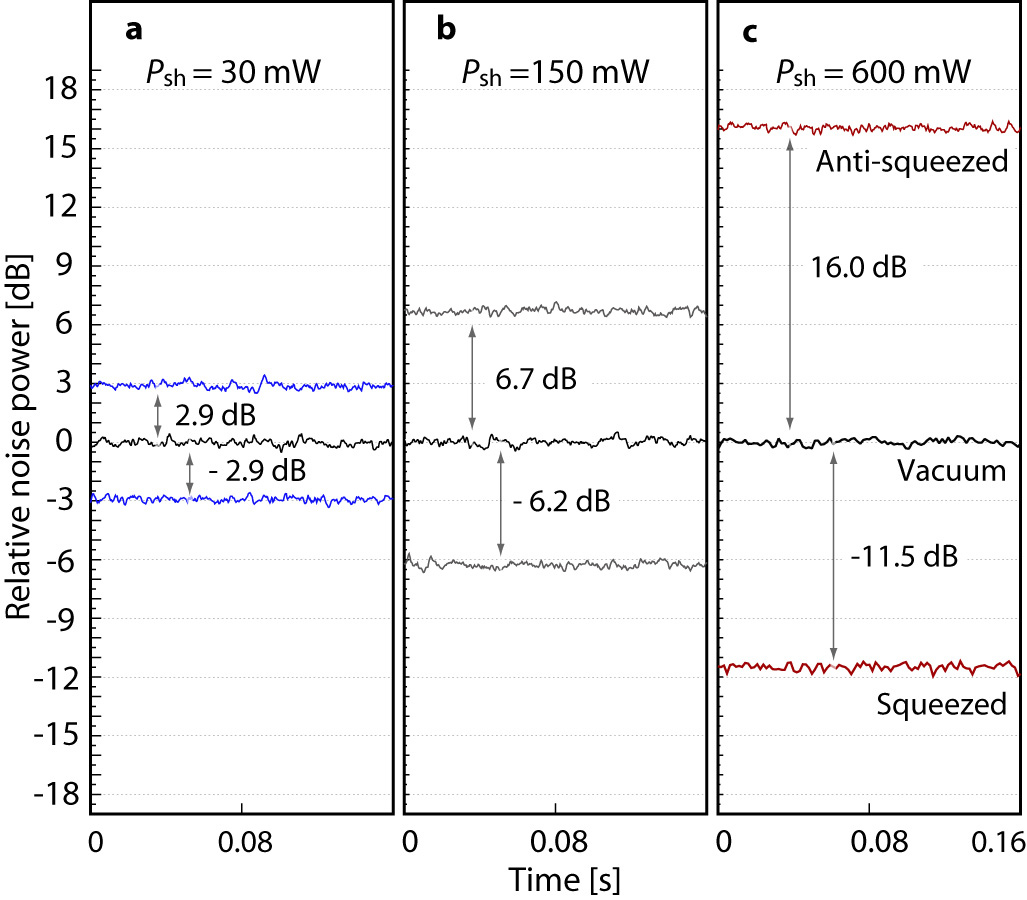}}
\vspace{0mm}
\caption{
Quadrature noise variances as measured by balanced homodyne detection. All traces were recorded at a Fourier frequency of $f$=5\,MHz, with a resolution bandwidth of $\Delta f=$100\,kHz and video bandwidth of 100\,Hz, and were normalized to the vacuum noise. The data still include electronic detector dark noise, hence all traces represent direct observations without any data post-processing. {\bf{(a)}}, Squeezed vacuum state that was generated with a second harmonic pump power of $P_{\mathrm{sh}}$=30\,mW. A nonclassical noise reduction of 2.9\,dB below vacuum noise was observed, with the respective anti-squeezing of 2.9\,dB above shot noise. {\bf{(b)}}, Squeezed vacuum state generated with an increased pump power of $P_{\mathrm{sh}}$=150\,mW. Squeezing of 6.2\,dB was measured with 6.7\,dB of anti-squeezing. {\bf{(c)}}, Driving the OPO at maximum pump power $P_{\mathrm{sh}}$=600\,mW yielded squeezing and anti-squeezing of 11.5\,dB and 16.0\,dB respectively. The subtraction of the detector dark noise would further increase the value of maximum squeezing to 11.6\,dB.
}
 \label{squeezing}
\end{figure}

The traces in Fig.~\ref{squeezing}(a)-(c) represent three different squeezed vacuum states from our OPO source in terms of pairs of squeezed (bottom trace) and anti-squeezed (top trace) vacuum normalized quadrature variances $V_1=\Delta^2 \hat X_1 / \Delta^2 \hat X_{1,\mathrm{vac}}$ and $V_2=\Delta^2 \hat X_2 / \Delta^2 \hat X_{2,\mathrm{vac}}$, respectively. 
For a pump power of 30\,mW (Fig.~\ref{squeezing}(a)) the squeezed variance ($V_1$) was 2.9($\pm$0.1)\,dB below and the anti-squeezed variance ($V_2$) 2.9($\pm$0.1)\,dB above the vacuum noise level. Within the measurement error bars of $\pm$0.1\,dB the state shows minimum uncertainty and therefore perfect purity. Upon driving the OPO with 150\,mW pump power (Fig.~\ref{squeezing}(b)) we observed normalized variances of -6.2($\pm$0.1)\,dB and 6.7($\pm$0.1)\,dB, respectively. The strongest and unprecedented squeezing level was measured with a pump power of approximately 600\,mW (Fig.~\ref{squeezing}(c)). With this setting 16.0($\pm$0.1)\,dB of anti-squeezing and a noise reduction of 11.5($\pm$0.1)\,dB below the vacuum noise level was achieved. \par
In practice, squeezed vacuum states generally cannot have perfect purity due to both residual optical loss and residual phase noise. The former results in an admixture of the vacuum state and the latter results in a mixture of phase space rotated states \cite{Taken07,FHDFS06}. 
Here, we make the attempt to describe all the three squeezed states measured by a fixed amount of optical loss assuming zero phase noise. 
This approach is reasonable for the following reasons, i) previous investigations have shown that our set-up is stable and phase noise is low \cite{Vahlb08}, ii) in contrast to squeezing in the audio-band \cite{Vahlb07NJP}, at MHz Fourier frequencies no displaced or thermal states are present that could mix into the state, and iii) no evidence was found for a non-Gaussian statistic in our measured data. All 3 pairs of variances observed can then be consistently calculated using $V_i = \eta\gamma V'_i + (1-\eta\gamma)$, where $V'_1=1/V'_2$ represent the pure state variances before any decoherence has occurred. Here, $(1-\eta\gamma)$ represents the overall contribution of the vacuum state, in power. The detection efficiency $\eta$ accounts for all losses outside the squeezed light source and $\gamma$ is the OPO cavity escape efficiency accounting for all losses inside the squeezed light source. We find that all of the measured variances (shown in Fig.~\ref{squeezing}) can indeed be modelled by a vacuum noise contribution as low as $(1-\eta\gamma)=(4.8\pm0.2)$\,\%, independent of the pump power used. This result supports our assumption that phase noise is not significant in our set-up.
The loss value of 4.8\% is consistent with an independent loss analysis of all the optical components used in our set-up, however, an independent measurement of the quantum efficiency of the photo diodes of the homodyne detector typically has large uncertainties, and we estimate this value to ($98\pm2$)\%. 
Such a low optical loss should enable the observation of more than 13\,dB of squeezing, however, pump powers above 600\,mW stably controlled and within a clean fundamental mode were not available in our experiment. 
\begin{figure}[b]
\centerline{\includegraphics[width=0.5\textwidth]{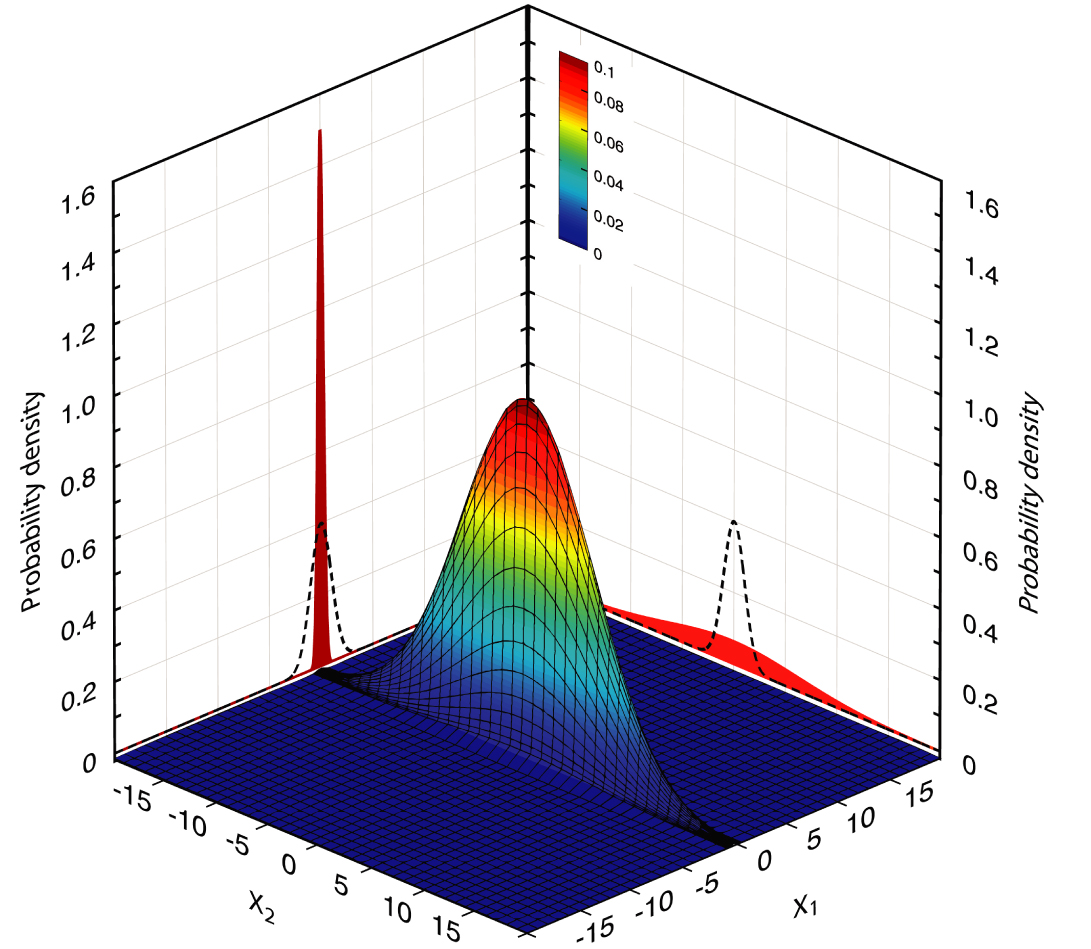}}
\vspace{0mm}
\caption{
Wigner function of the squeezed vacuum state produced by our OPO. The projections (filled curves) onto the two quadratures yield the gaussian probability distributions with variances of -11.5\,dB and 16\,dB relative to the projections belonging to a pure vacuum state (dotted curves).
}
 \label{wignerfct}
\end{figure}
Figure~\ref{wignerfct} presents the Wigner function \cite{Wigner32,Bertrand87,Vogel89} of the squeezed vacuum state with variances of -11.5\,dB and 16\,dB with respect to the vacuum noise variance. The Wigner function represents one of several possible phase-space probability distributions for quantum states. Since it is Gaussian for vacuum-squeezed states, it is particularly easy to obtain simply by measuring its
second moments. The entire statistical properties of the state can than be
summarized in a finite dimensional space, namely, by the covariance matrix. Hence, for squeezed vacuum states mixed with the ordinary vacuum state only the variances of the squeezed and anti-squeezed quadrature are required to calculate the Wigner function which is then given by
\begin{equation}\nonumber
W(X_1,X_2) = \frac{(4\pi)^{-1}}{\sqrt{\Delta^2 \hat X_1 \Delta^2 \hat X_2}} \exp\!{\left( \frac {-\frac{1}{2} X_{1}^{2}}{\Delta^2 \hat X_1}+\frac {-\frac{1}{2}X_{2}^{2}}{ \Delta^2 \hat X_2} \right)}\;,
\end{equation}
where $\Delta^2 \hat X_i$ are again the variances of the quadratures $X_{1}$ and $X_{2}$, respectively, with the vacuum noise variance of 1/4. The vertical axis in Fig.\ref{wignerfct} represents the quasi-probability density of the quadrature phase space. The Wigner function is a phase-space \textit{quasi}-probability distribution, since the quadratures cannot simultaneously have precisely defined values. What \emph{can} be precisely measured is one projection of the Wigner function onto the axis of the phase space at a time. The projections are the probability distributions of the particular quadrature, and here, have a Gaussian shape with squeezed or anti-squeezed variances with respect to the vacuum noise variance (dotted curves). Contour lines of the Wigner function have an elliptical shape and are thus squeezed in comparison with the circular contour lines of the vacuum state (not shown in Fig.~\ref{wignerfct}).
\subsection{Squeezed vacuum states in the photon number basis}
\begin{figure}
  \centerline{\includegraphics[width=0.5\textwidth]{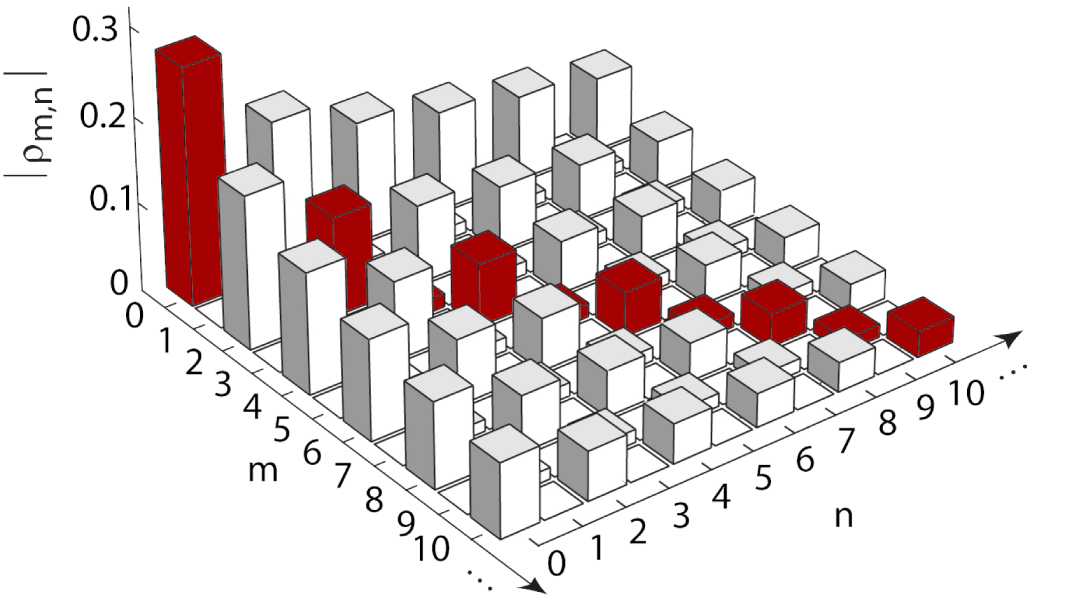}}
  \vspace{0mm}
\caption{
Density matrix of the squeezed state of Fig.~\ref{squeezing}(c). Plot of the absolute values of the density matrix elements with photon numbers up to 10 for the -11.5\,dB squeezed and 16\,dB anti-squeezed state. The red diagonal elements give the probabilities of finding 0,1,2\dots n photons. Due to the state's high purity, the probabilities for finding an odd number of photons is significantly suppressed.
}
 \label{dM}
\end{figure}

All photons that are present in a squeezed vacuum state produced by OPO stem from parametric down conversion of second harmonic pump photons. This is also true for a mixture of a squeezed vacuum state with the ordinary vacuum state. For strongly squeezed states not just the detection of photon pairs but also of higher-order photon pairs is expected. However, the overall loss of photons must be sufficiently low to enable such an observation. 
Since the Wigner function represents the full quantum information of a state it can be used to calculate the state's density matrix $\rho_{n,m}$ in the photon number (Fock) basis, which is another complete description of the state. The photon number distribution P(n) can be obtained from the diagonal elements of the density matrix. Figure~\ref{dM} shows a plot of the first matrix elements (absolute values) up to photon numbers $n,m=10$ of the observed squeezed state with noise variances of -11.5\,dB and 16\,dB relative to the vacuum noise. In order to calculate the density matrix \cite{GerryKnight,Marek07} photon numbers of up to $n=m=170$ were considered, which ensured convergence of all the matrix elements.
Similarly, the density matrices for the additional two states have also been generated. Their diagonal elements, i.e. the photon number distributions of the states, are depicted in Fig.~\ref{photonnumber}. The formula used and the numerical values of the density matrix elements are given in the supplementary information. For the 11.5\,dB squeezed state (red) a strong odd/even photon number oscillation is visible. Even the probability of detecting $n=10$ photons is significantly higher than the probability of detecting $n=9$ photons. This oscillation further continues for greater photon numbers but is not shown here. For our squeezed states with less squeezing the oscillation flattens out at lower photon numbers, but the odd/even oscillation is even more pronounced at lower photon numbers. For the 2.9\,dB squeezed state the probability of finding $n=1$ photon is as small as 0.5\%.     
\begin{figure}[b]
\includegraphics[width=0.5\textwidth]{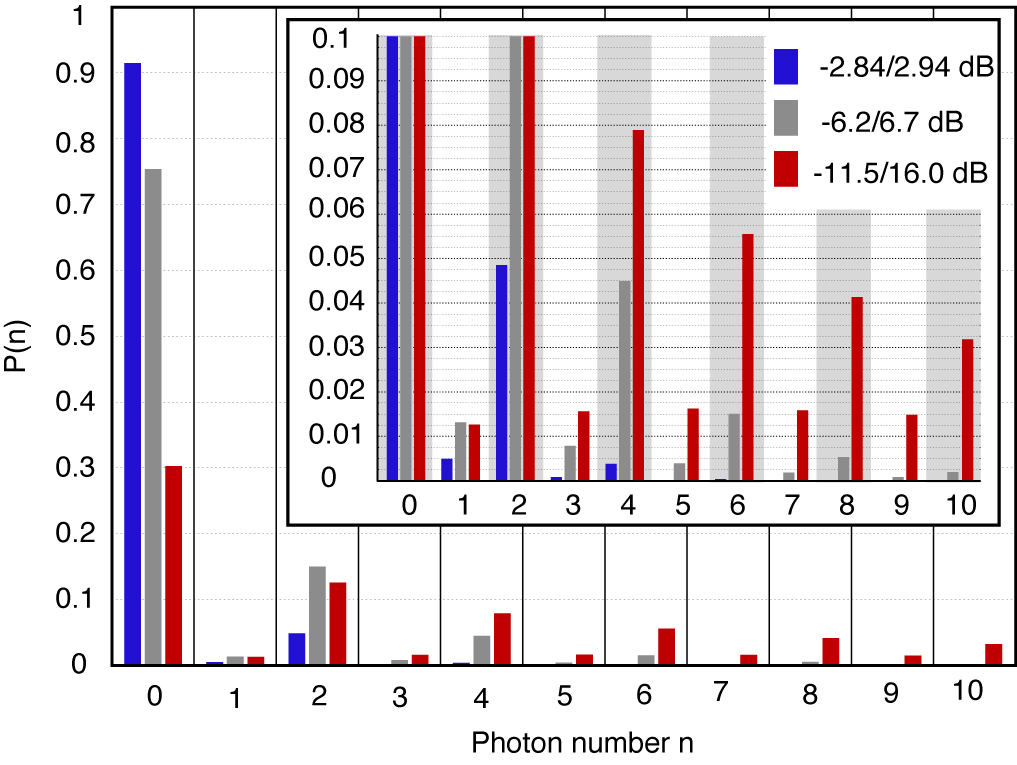}
\vspace{0mm}
\caption{
Photon number oscillation. The photon number distribution of the three squeezed vacuum states that were generated in our setup
clearly show oscillatory behavior. With increasing squeezing the number of higher order photon pairs grows. For clarity the inset shows a zoomed view restricted to probabilities up to 0.1. 
}
 \label{photonnumber}
\end{figure}
Note that for \textit{displaced} squeezed states additional photons are added to the state and the odd photon numbers can not all have zero probability, even if these state were pure. However, \textit{weakly} displaced squeezed states show photon number oscillations with larger periods. Such states were originally theoretically analyzed in \cite{SWh87}.\\
The photon numbers calculated here are detected per bandwidth in Hertz and per detection time interval in seconds. Obviously, the higher the squeezing bandwidth the higher the number of photons per second.
\subsection{Observation of high-bandwidth squeezing} \label{subsectionSpectrum} 
For applications in high speed quantum communication and information processing high photon counting rates or, equivalently, fields with squeezing over a broad Fourier spectrum are essential. The half width half maximum (HWHM) \textit{squeezing bandwidth} can be defined as the spectral width over which the squeezed variance increases from its lowest variance at small Fourier frequencies $V_1$ to $V_1+0.5\cdot(1-V_1)$. Here the variances are again normalized to unity vacuum noise. Although continuous variable quantum states are often quoted to provide the possibility of high bandwidths, only modest values of the order of a few tens of MHz have been realized, e.g. in \cite{Breitenbach95,Chelk07}.
\begin{figure}[b]
 \centerline{\includegraphics[width=0.5\textwidth]{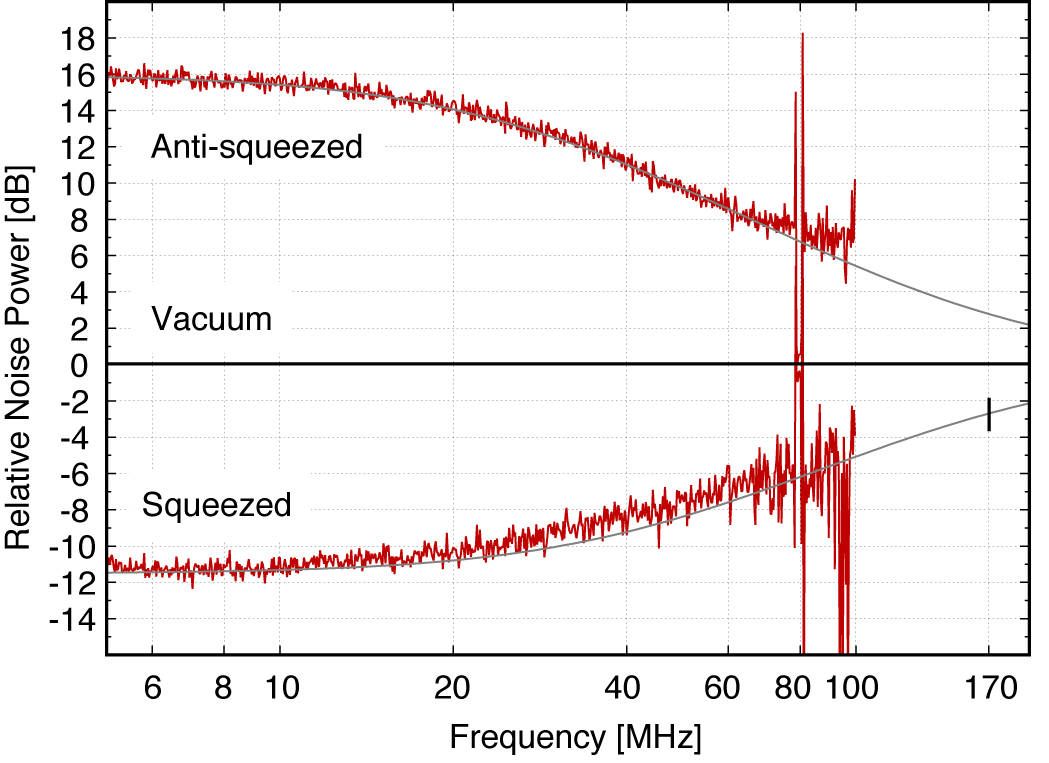}}
  \vspace{0mm}
\caption{
High bandwidth squeezing spectrum. Squeezing (bottom trace) and anti-squeezing (top trace) are shown relative to the vacuum noise variance. The measurements were performed from 5\,MHz to 100\,MHz. This was limited by the fact that the dark noise clearance of the homodyne detector was too low towards higher frequencies. The traces are averages of 3 measurements each done with a resolution bandwidth of 1\,MHz and video bandwidth of 3\,kHz. The data was fitted to a model and revealed a squeezing bandwidth of 170\,MHz (thin line).
}
 \label{broadbandsqueezing}
\end{figure}
Figure~\ref{broadbandsqueezing} shows the spectral analysis of the strongly squeezed field from our source with a HWHM squeezing bandwidth of 170\,MHz. In contrast to Fig.~\ref{squeezing} here the electronic dark noise of our homodyne detector was subtracted. At $\sim$100\,MHz the dark noise reached the measured quantum noise. At frequencies below 5\,MHz the measurement was limited by the resolution bandwidth chosen. Since our set-up did not use control beams which might introduce technical noise at low frequencies, a white squeezing spectrum can be expected for lower frequencies if parasitic interferences with frequency shifted scattered photons are sufficiently suppressed as demonstrated in \cite{Vahlb06PRL,Vahlb07NJP}.\\
The spectrum of the squeezed and anti-squeezed quadrature variances ($V_{1}$ and $V_{2}$, respectively) for the below-threshold OPO can be calculated by \cite{Taken07}
\begin{equation}\label{sqzspec}
    V_{1,2} = 1\pm \eta \gamma \frac{4 \sqrt{P_{\textrm{sh}}/P_{\textrm{th}}}}{(1\mp \sqrt{P_{\textrm{sh}}/P_{\textrm{th}}})^2+4K^2}\ ,
\end{equation}
where $P_{532}$ is the power of the second harmonic pump and $P_{th}$ is the pump power needed to reach the OPO threshold. The detection efficiency $\eta$ is calculated as the product of the propagation efficiency, the square of the homodyne visibility, and the quantum efficiency of the homodyne detector photo diodes. The OPO escape efficiency $\gamma=T/(T+L)$ is determined by the transmittance $T$ of the out coupling mirror and the intra-cavity round trip loss $L$. The dimensionless parameter  $K=2\pi f/\kappa$ is the ratio of the Fourier frequency $f$ and the decay rate $\kappa=(T+L)c/l$ of the cavity with round trip length $l$.
The theoretical curves in Fig.~\ref{broadbandsqueezing} were modeled using equation~(\ref{sqzspec}), with the best agreement obtained upon assumption of the following values: a pump beam at 53.5\% of the threshold power, the output coupler transmittance of 12\%, the round-trip loss within the crystal of approximately 0.1\%, and a total detection efficiency $\eta \gamma$=95.2\% in agreement with our analysis presented in section~\ref{subsection1}. The squeezing bandwidth of our source is given by the Fourier frequency where the squeezed variance has increased to -2.7\,dB. 
According to the theoretical curve in Fig.~\ref{broadbandsqueezing} this bandwidth is 170\,MHz. This result can be used to calculate the rate of down-converted photons from a full longitudinal mode of our source. To this end, we split the half free spectral range of our source of 5.5\,GHz into 55000 frequency bins of 100\,kHz each. For every bin $i$, the photon number distribution $P_i(n)$ was calculated. By summing these distributions, we find that the spectrally weighted representative photon number distribution of the central OPO cavity mode around half the pump light frequency  $P_{f}(n) = 1/55000\cdot\sum_i P_i(n)$. This distribution can now be used to calculate the mean number of photons per second and bandwidth averaged over the band of 5.5\,GHz. Multiplying with 5.5\,GHz provides the rate of down-converted photons from this mode of as high as $2.79\cdot10^8$s$^{-1}$, corresponding to a power of 52\,pW.\\
Another relevant value is given by the mean number of correlated photons in the case of a single photon detection. For a parametric down-conversion source operated far below its threshold and with zero photon loss this number is two because the detection of a single photon proves the presence of, in total, two photons. In order to calculate this number for our source we set $P_{f}(0) = 0$ and calculate the mean value of the renormalized remaining distribution. For the 11.5\,dB squeezed state presented in Fig.\ref{broadbandsqueezing} we obtain $\langle n \rangle_{f}$ = 5.93. Implying that whenever a single photon detector clicks (possibly without any photon number resolution) this event proves the presence of, on average, almost 6 quantum correlated photons, including the detected photon(s).
%
\section{Discussion}   
In order to observe strong squeezing levels, we operated an optical parametric oscillator comparatively close to its threshold. In this regime not just single photon pairs, but a significant number of multiple or \textit{higher order} photon pairs were produced.
This contrasts typical parametric down-conversion experiments in the photon counting regime where the probability of observing more than two quantum correlated photons is negligible. The density matrix in the photon number basis of our -11.5\,dB squeezed state not only reveals strong photon number oscillations but also quantum correlations between photons in the following way: Whenever a photon was photo-electrically detected by the homodyne detector, there was an almost unity probability that an additional, odd number of photons were physically realized and were detected by the same detector. In photon counting experiments our squeezed state might therefore be used to overcome the problem of insufficient photon counter efficiencies. 
The unprecedented quality of our squeezed states will also enable a great step forward in many applications currently under discussion. In high power laser interferometers for gravitational wave detection \cite{WPAUGZ04,VCHFDS05,Goda08NatPhys} the application of an -11.5\,dB squeezed state reduces the shot noise equivalent to a laser power increase by a factor 14. The interference of two such states on a 50/50 beamsplitter would produce strong Einstein-Podolsky-Rosen entanglement \cite{FSBFKP98,BTBSRBSL03,James07} and its high bandwidth may result in the ultra-fast generation of quantum keys \cite{Cerf01}. Our squeezed light source may also find applications in future hybrid quantum information networks. Its well-defined fundamental Gaussian spatial mode yielded a high fringe contrast with a spatially filtered local oscillator of 99.8\% and enables efficient coupling to other systems, for example to atoms coupled to optical cavities~\cite{Lukin00,Simon07,Herskind09}, in order to realize quantum interfaces of light and matter.

\section*{ACKNOWLEDGEMENTS}
We would like to thank B. Hage, J. DiGuglielmo, A. Th\"uring, C. Messenger and J. Fiur\'{a}\v{s}ek for many helpful discussions. This work has been supported by the Deutsche Forschungsgemeinschaft (DFG) through Sonderforschungsbereich 407 and the Centre for Quantum Engineering and Space-Time Research QUEST.

\section*{APPENDIX}
In order to calculate the density matrix elements $\rho_{m,n}$ we followed the prescription given in~\cite{Marek07}, whereby the entries can be obtained using the following equation:
\begin{equation}
\rho_{m,\,n} = \sqrt{\frac{m!n!}{\tilde{V}_x \tilde{V}_p}}\sum_a
\frac{(-T)^{\left[(m-n)/2\right]+2\alpha}U^{n-2\alpha}}{a!(n-2a)!\left(a +
\frac{m-n}{2}\right)!},
\end{equation}
where $\tilde{V}_x = V_x + 1/2$, $\tilde{V}_p = V_p + 1/2$, $U = 1 -
\frac{1}{2 \tilde{V}_x} - \frac{1}{2 \tilde{V}_p}$ and $T = \frac{1}{4
\tilde{V}_x} - \frac{1}{4 \tilde{V}_p}$. 
Note, that $V_x$ and $V_p$ correspond to the linear quadrature variances of the quantum state as introduced in equation~(\ref{sqzspec}), but with the vacuum variance set to 1/2. 
This expression for the density matrix in the Fock state basis can be derived from the covariance matrix by first noting that the Husimi Q-function in the coherent state basis is the generating function for the density matrix in the Fock state basis. The covariance matrix needs to be transformed from the quadrature basis into the coherent basis to obtain the above formula.\\
We calculated $\rho_{m,n}$ for the three states in Fig.\ref{squeezing}a-c.
For the first state (Fig.\ref{squeezing}a) we chose vacuum normalized variances of -2.84\,dB and 2.94\,dB being within our experimental error bars. These numbers correspond to an originally pure state with 3.05\,dB squeezing and 3.05\,dB anti-squeezing that was degraded by 4.8\% of optical loss. For the remaining two states we chose the variances according to the values in Fig.\ref{squeezing}b (-6.2\,dB and 6.7\,dB) and Fig.\ref{squeezing}c (-11.5\,dB and 16.0\,dB). The calculation was done with photon numbers up to 170 to ensure the convergence of all entries, but the matrices presented below were truncated at n=m=10. For clarity the diagonal elements representing the photon number distribution (as shown in Fig.\ref{photonnumber}) are marked in red.
\onecolumngrid
\setcounter{MaxMatrixCols}{12}
\begin{center}
$
\rho_{-2.84\textrm{dB}} =\left[
\begin{matrix}
	\tcr{.9416}&0&-.2137&0&.0594&0&-.0174&0&.0052&0&-.0016\\
	0&\tcr{.0049}&0&-.0019&0&.0007&0&-.0002&0&.0001&0\\
	-.2137&0&\tcr{.0485}&0&-.0135&0&.0040&0&-.0012&0&.0004\\
	0&-.0019&0&\tcr{.0008}&0&-.0003&0&.0001&0&-0&0\\
	.0594&0&-.0135&0&\tcr{.0038}&0&-.0011&0&.0003&0&-.0001\\
	0&.0007&0&-.0003&0&\tcr{.0001}&0&-0&0&0&0\\
	-.0174&0&.0040&0&-.0011&0&\tcr{.0003}&0&-.0001&0&0\\
	0&-.0002&0&.0001&0&-0&0&\tcr{0}&0&0&0\\
	.0052&0&-.0012&0&.0003&0&-.0001&0&\tcr{0}&0&0\\
	0&.0001&0&-0&0&0&0&0&0&\tcr{0}&0\\
	-.0016&0&.0004&0&-.0001&0&0&0&0&0&\tcr{0}\\
\end{matrix}
\right] $.
\end{center}
\setcounter{MaxMatrixCols}{10}

\setcounter{MaxMatrixCols}{12}
\begin{center}
$
 \rho_{-6.2\textrm{dB}} =\left[
\begin{matrix}
\tcr{.7538}&0&-.3360&0&.1834&0&-.1056&0&.0622&0&-.0372    \\
0&\tcr{.0131}&0&-.0101&0&.0071&0&-.0048&0&.0032&0\\
-.3360&0& \tcr{.1500}&0&-.0820&0&.0473&0&-.0279&0&.0167\\
0&-.0101&0& \tcr{.0078}&0&-.0055&0&.0037&0&-.0025&0\\
.1834&0&-.0820&0& \tcr{.0449}&0&-.0259&0&.0153&0&-.0092\\
0&.0071&0&-.0055&0&\tcr{.0039}&0&-.0026&0&.0018&0\\
-.1056&0&.0473&0&-.0259&0& \tcr{.0150}&0&-.0089&0&.0053\\
0&-.0048&0&.0037&0&-.0026&0&\tcr{.0018}&0&-.0012&0\\
.0622&0&-.0279&0&.0153&0&-.0089&0&\tcr{.0053}&0&-.0032\\
0&.0032&0&-.0025&0&.0018&0&-.0012&0&\tcr{.0008}&0\\
-.0372&0&.0167&0&-.0092&0&.0053&0&-.0032&0&\tcr{.0019}\\
\end{matrix}
\right] $.
\end{center}
\setcounter{MaxMatrixCols}{10}

\setcounter{MaxMatrixCols}{12}
\begin{center}
$
 \rho_{-11.5\textrm{dB}} =\left[
\begin{matrix}
\tcr{.3026}&0&-.1946&0&.1532&0&-.1272&0&.1082&0&-.0933 \\
0&\tcr{.0126}&0&-.0140&0&.0143&0&-.0140&0&.0135&0 \\
-.1946&0&\tcr{.1256}&0&-.0993&0&.0828&0&-.0707&0&.0613 \\
0&-.0140&0&\tcr{.0156}&0&-.0159&0&.0157&0&-.0151&0 \\
.1532&0&-.0993&0&\tcr{.0789}&0&-.0660&0&.0566&0&-.0493 \\
0&.0143&0&-.0159&0&\tcr{.0162}&0&-.0160&0&.0155&0 \\
-.1272&0&.0828&0&-.0660&0&\tcr{.0555}&0&-.0478&0&.0417 \\
0&-.0140&0&.0157&0&-.0160&0&\tcr{.0158}&0&-.0153&0 \\
.1082&0&-.0707&0&.0566&0&-.0478&0&\tcr{.0413}&0&-.0362 \\
0&.0135&0&-.0151&0&.0155&0&-.0153&0&\tcr{.0148}&0 \\
-.0933&0&.0613&0&-.0493&0&.0417&0&-.0362&0&\tcr{.0318} \\
\end{matrix}
\right] $.
\end{center}
\setcounter{MaxMatrixCols}{10}

\twocolumngrid


\end{document}